# Error Threshold of Fully Random Eigen Model

LI Duo-Fang[1], CAO Tian-Guang[1], GENG Jin-Peng[1],QIAO Li-Hua[2],

GU Jian-Zhong[3], ZHAN Yong[1]*

[1] *Institute of Biophysics, School of Sciences, Hebei University of Technology, Tianjin 300401, China*

[2] *School of Basic Medicine, Heibei Medical University, Shijiazhuang 050017, China*

[3] *Nuclear Physics Institute, China Institute of Atomic Energy, Beijing 102413, China*

*Species evolution is essentially a random process of interaction between biological populations and their environments. As a result, some physical parameters in evolution models are subject to statistical fluctuations. In this paper, two important parameters in the Eigen model, the fitness and mutation rate, are treated as Gaussian distributed random variables simultaneously to examine the property of the error threshold. Numerical simulation results show that the error threshold in the fully random model appears as a crossover region instead of a phase transition point, and as the fluctuation strength increases the crossover region becomes smoother and smoother. Furthermore, it is shown that the randomization of the mutation rate plays a dominant role in changing the error threshold in the fully random model, which is consistent with the existing experimental data. The implication of the threshold change due to the randomization for antiviral strategies is discussed.*

Darwin's theory of evolution, with a profound impact on biology, proposes that the selection and mutation are two main driving forces of biology evolution. In 1971, chemist M. Eigen introduced the selection and mutation into lifeless chemical reaction systems and established a mathematical model of species evolution, known as the Eigen model [1]. The Eigen model has two important predictions: the quasi-species and error threshold. The quasi-species means the distribution of some mutant sequences localized around the master sequence. The error threshold is a point above which there is no genetic information in the macromolecules and all of the sequences are randomly distributed in the sequence space [2, 3]. The characteristics of the quasi-species and error threshold have been confirmed by many experiments [4-8]. The studies on the Eigen model have some important implications for virus evolution and cancer treatment.

In the past four decades, the studies of the Eigen model have focused on its dynamic solution and equilibrium characteristic with different fitness landscapes [9-13]. Species evolution in nature, however, is always affected by various random factors, such as genetic mutations and environment fluctuations [14, 15]. As a consequence, the important physical parameters appearing in species evolution models are also subject to these kinds of random factors and should then be stochastic. Therefore, it is reasonable that these parameters should be considered as random variables. In fact, the inherent randomicity of physical parameters in species evolution models has been confirmed by virus experiments [7, 16, 17].

In previous studies, the fitness and mutation rate in the Eigen model were treated as Gaussian distributed random variables, separately [18, 19]. In this paper, the fitness and mutation rate in the Eigen model are replaced simultaneously by Gaussian distributed random variables in order to investigate the change of the error threshold due to the cooperative fluctuations of the fitness and mutation rate.

The Eigen model describes the evolution of self-replicating molecules which represent different macromolecules with specific genetic information. These macromolecules, as sequences, have a fixed length $N$. If one only considers the purine and pyramidine in a RNA or DNA molecule, each site of the molecule is a binary variable, and the number of all possible sequences is $2^N$. To simplify the model, the Hamming distance is introduced to classify the possible sequences, which is defined as the number of different bases between arbitrary two sequences. The sequences with the same Hamming distance from the master sequence can be combined into a class, and one has therefore N+1 classes. The mathematical equation of the Eigen model is given as follow:

$$\frac{dx_i}{dt} = \sum_{j=0}^{N} f_j q_{ij} x_j - \phi(\vec{x}) x_i \quad (1)$$

Here $x_i$ is the relative concentration of the $i$-th class $I_i$ in the whole population. $f_0$ and $f_i$ are respectively the fitness of the master class $I_0$ and the fitness of the mutant class $I_i$. Single-peaked fitness landscape [20] is assumed here. So we take $f_0 = A_0$ ， $f_i = A_i < A_0$ $(i \neq 0)$ and $A_0$ and $A_i$ are constants. $q_{ij}$ stands for the transition probability from class $I_j$ to class $I_i$. The dilution flux $\phi$ is introduced to keep overall of molecules constant, which satisfies the condition of $\phi(\dot{x}) = \sum_{i=0}^{N} f_i x_i$. Equation (1) is obviously a nonlinear differential equation. By some variable transformation, a linear differential equation [21] of the Eigen model can be obtained. The normalized eigenvector corresponding to the largest eigenvalue of the rate matrix $W_{ij} = f_j \cdot q_{ij}$ gives the relative concentrations in their stationary states. $q_{ij}$ can be written as:

$$\mathrm{q}_{ij} = \sum_{l=l_{\min}}^{l_{\max}} \binom{j}{l} \binom{N-j}{i-l} q^{N-j-i+2l} (1-q)^{j+i-2l} \quad (2)$$

Here $q$ denotes copying fidelity of each site of a sequence. $u = 1 - q$ is the mutation rate. $l_{\min} = \max\{0, j+i-N\}$, $l_{\max} = \min\{j, i\}$.

To investigate stochastic characteristics of the Eigen model, the fitness and mutation rate are treated simultaneously as Gaussian distributed random variables. Their probability density distributions are:

$$\mathrm{P}(\mathrm{y}_i) = \frac{1}{\sqrt{2\pi\omega^2}} e^{\frac{-(y_i - \bar{y})^2}{2\omega^2}} \quad (3)$$

where $y_i$ is a variable. $\bar{\mathrm{y}}$ and $\omega^2$ denote the averaged value and variance of the random variable. In the case of the random fitness, the averaged values of the master class and mutant classes are $A_0$ and $A_i$ respectively. In the case of the random mutation rate, the averaged value of the mutation rate is $u = 1 - q$. The fluctuation strength of the random variable is given by $d = \omega / \bar{y}$.

Then the fitness and mutation rate in the deterministic Eigen model are replaced by random sampling values. It is called fully random Eigen model. Thereby the random model would produce an ensemble of quasi-species population. The features of the error threshold in the random model are studied by analyzing the statistical properties given by the ensemble.

In the studies of the random model, the quality of the random number could affect the reliability of the results. Therefore, let us test the reliability of the Gaussian distributed variables appearing in the fully random Eigen model. By 10000 random sampling, the distributions with $A_0 = 10$, $A_i = 1$, and $u = 0.1$ are obtained by the frequency statistic and Gaussian fitting. The results for the fluctuation strength $d$=0.1 are shown in Fig. 1. One can see that both the fitness and mutation rate are nicely fitted by the Gaussian distributions, and the goodness of fit ( $R^2$ ) are all above 0.99. For all values of the fluctuation strength in the simulation (0-0.20), the fitting results abide basically the Gaussian distributions. The test of the Gaussian distribution demonstrates the reliability of the random variables used in our calculations. When the value of the fluctuation strength is greater than 0.20, the small probability sampling occurs inevitably, resulting in the singularity [3]. Thus, there should be the upper limit of the fluctuation strength of the random variables in simulations.

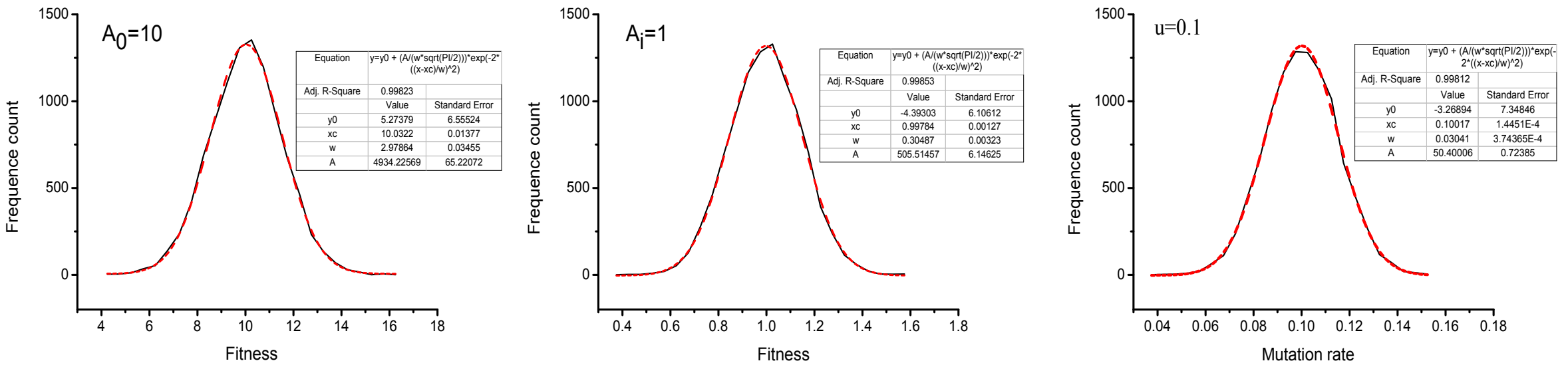

Fig. 1. Gaussian test for the random variables in the cases of $A_0 = 10$, $A_i = 1$, and $u = 0.1$, with the fluctuation strength $d = 0.1$. The black lines are numerical simulation results and the red dotted lines are given by the formulas.

To analyze the nature of the error threshold in the random model, the above Gaussian distributed random variables are used. The ensemble of the relative concentration in the stationary state for each class is obtained. Fig. 2 shows the distributions of the averaged relative concentrations in the deterministic model ( $d = 0$ ) and the random model ( $d = 0.15$ ), where 0 represents the master sequence, and 1, 2, 3… represent different mutant classes. As is shown in Fig. 2A, the error threshold is at 0.112, with the feature of a sharp phase transition in the deterministic model. Below the error threshold, the quasi-species distribution appears, and some mutation sequences are localized around the master sequence. And above the error threshold, all sequences have the same negligible concentration. It can be seen that the relative concentrations of the complementary classes ($I_i$ and $I_j$, $i + j = N$ ) are identical beyond the error threshold.

In Fig. 2B, compared with the situation in the deterministic model, the error threshold in the fully random Eigen model becomes smooth, like a crossover region, which is consistent with previous studies [18, 19]. The error threshold is located in the range of 0.112-0.190 when the fluctuation strength is taken to be $d = 0.15$. This indicates that the upper limit of the crossover region should be considered when dealing with practical problems. Although the error threshold changes obviously in the fully random model, the concentrations in the other regions are roughly consistent with those in the deterministic model, suggesting that they are relative steady against the fluctuation in the fitness and mutation rate.

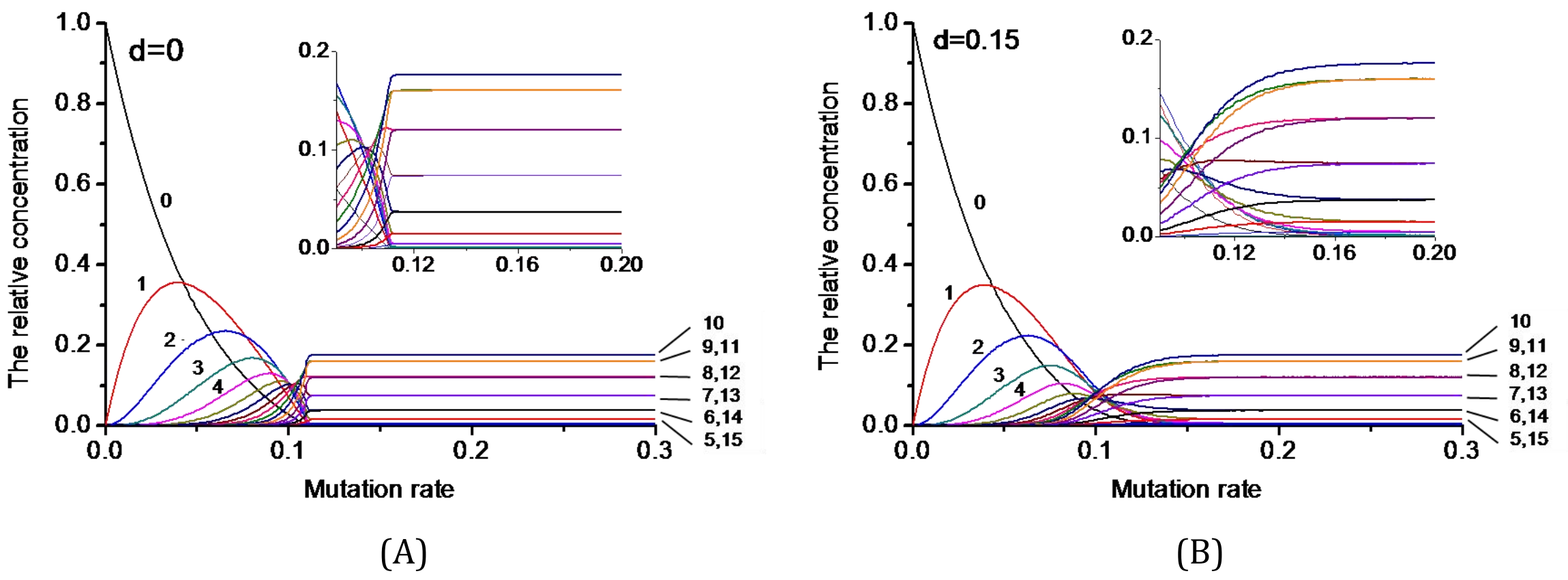

Fig. 2. The calculated relative concentrations of quasi-species classes in their stationary states under the values of the fluctuation strength $d = 0$ and $d = 0.15$ with $N = 20$, $A_0 = 10$, $A_i = 1$, $A_0/A_i = 10$, $\omega_0 = 1.5$, $\omega_1 = 0.15$, and the number of random samples $n = 10000$.

The range of the error threshold in fully random model is defined in order to quantitatively examine the relationship between the change of the error threshold and the fluctuation strength of the random variables. The starting point of the error threshold range is the phase point given by the deterministic model. The end point is the position where the relative difference between the relative concentrations of two complementary classes is less than 0.01, and the relative difference *c* is given by

$$c = \frac{x_i - x_j}{(x_i + x_j)/2} \tag{4}$$

For different complementary classes, their relative concentrations behave in a similar manner in the crossover region. According to the above definition of the error crossover region, the ranges and widths of the crossover region for different fluctuation strength values are determined and shown in Table 1. It is seen that the crossover region becomes wider and wider as the fluctuation strength increases. The relationship between the width of the crossover region and the fluctuation strength *d* is nonlinear.

Table 1. The feature of the error threshold with different values of the fluctuation strength (*d*) in the case of the fully random model.

| Fluctuation strength (*d*) | Region of error threshold | Width of error threshold |
|---|---|---|
| 0 | 0.112 | 0 |
| 0.05 | 0.112-0.127 | 0.015 |
| 0.10 | 0.112-0.149 | 0.037 |
| 0.15 | 0.112-0.182 | 0.070 |
| 0.20 | 0.112-0.229 | 0.117 |

In the biology evolution, the fitness and mutation rate are interconnected and jointly drive organism evolution and development. The fitness and mutation rate are replaced separately by Gaussian distributed random variables, there is also a crossover region around the error threshold as demonstrated in references [18, 19]. In order to better understand the crossover region in the fully random Eigen model, let us compare our simulation results with those from the above references according to the crossover region defined in this article. In Figure 3, it is shown that the width of the crossover region changes with the fluctuation strength in the different three random simulations. It is worth noting that the upper limits of the fluctuation strength in single random physical parameter models are 0.25, and higher than that in the fully random model. The comparison analysis reveals that the change of the error threshold is least in the case of random fitness. The changes in the fully random model and in the case of random mutation rate are larger, while the difference between the two cases is rather small. This implies that the mutation rate fluctuation is crucial for quasi-species classes to pass through the crossover region and go extinction. In virus evolution experiments, it was found that increasing mutation rate usually renders virus more easily to fall into the error catastrophe than reducing fitness [7], which is similar to our results.

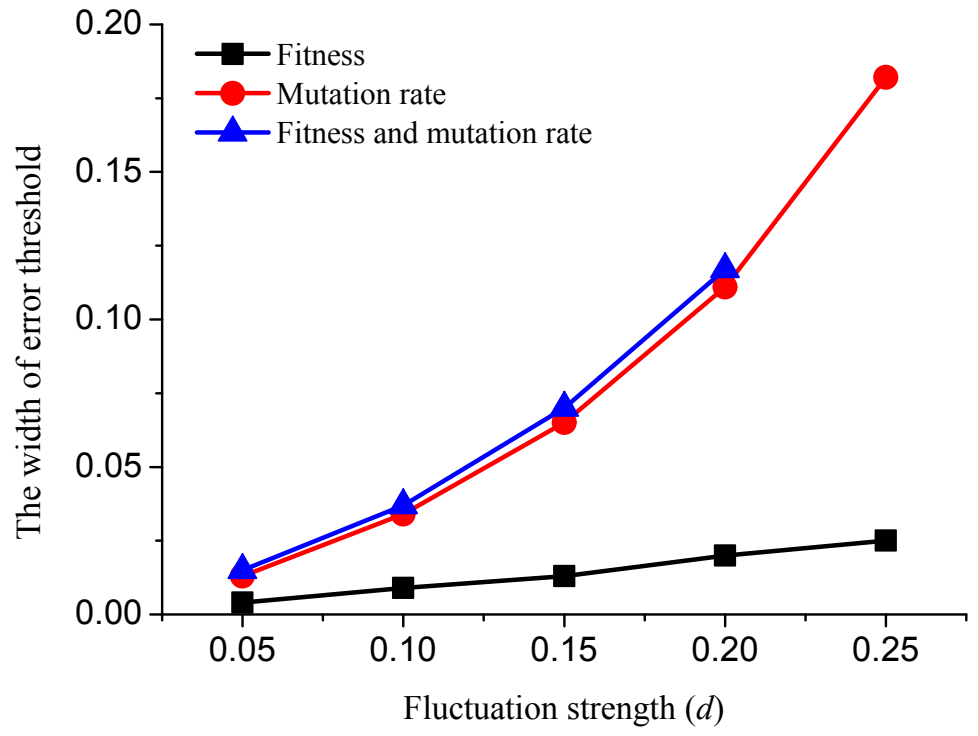

Fig. 3. The error crossover width versus the fluctuation strength in the cases of three different random simulations. The squares represent the results in the case of random fitness. The dots are obtained in the case of random mutation rate The triangles are the results in the case of the fully random model.

The most important contribution of the Eigen model is the discovery of quasi-species and error threshold, which has a very important significance for understanding the essences of nature, such as biological diversity, life origin and evolution. In particular, it is interesting that the character of the error threshold is very similar to the phase transition in physics. It was demonstrated that there existed quasi-species distributions and error thresholds of different virus systems by test tube experiments[4, 22] , which provided an enlightenment to develop new antiviral strategies. Recently, it was found that the error catastrophe phenomenon existed in the cancer cell evolution [23, 24]. Despite Eigen theory has made some achievements, there is still a long distance from the practical application. The researches of a stochastic Eigen model, in particular for the inherent randomicity of physical parameters, will render the Eigen theory more realistic for the description of species evolution and the design of antiviral strategies. The research on the randomness of biology processes has been quite active in both theoretical and experimental aspects [25-28].

In this paper, two important parameters, the fitness and mutation rate were randomized simultaneously. The results have suggested that as the fluctuation strength increases, the crossover region becomes smoother and smoother. Moreover, the comparison of the fully random model with the single random parameter models has indicated that the width of the crossover region in the fully random model is largest, and the mutation rate randomization plays a dominant role. Certainly, the fully random effects are related to the special form of the fitness and mutation rate rather than a simple linear superposition of them.

The above conclusions are especially significant for the viruses and pathogenic bacteria survival in complex hosting environments. The extension effect of the error threshold in random models could imply that to effectively drive viruses to go extinct, the upper limit of the crossover region should be reached.

## Acknowledgements

The authors are greatly indebted to Prof. Yizhong Zhuo for his valuable discussions and suggestions.